\documentclass[12pt]{iopart}
\usepackage[T1]{fontenc}
\usepackage[latin9]{inputenc}
\usepackage{amsbsy}
\usepackage{amstext}
\usepackage{esint}

\begin{document}

\title[AB effect and charge non-superselection]{AB effect and Aharonov-Susskind charge non-superselection }

\author{N. Erez}

%\maketitle

\address{Dept. of Particle Physics, Tel Aviv University, School of Physics and Astronomy, Tel Aviv University, 69978 Tel Aviv, Israel and Chemical Physics Department, Weizmann Institute of Science 76100 Rehovot, Israel}
\ead{nerez@weizmann.ac.il}

\begin{abstract}
We consider a particle in a coherent superposition of states with different electric charge moving in the vicinity of a magnetic flux. Formally, it should acquire a (gauge-dependent) AB relative phase between the charge states, even for an incomplete loop. If measureable, such a geometric, rather than topological, AB-phase would seem to break gauge invariance. Wick, Wightman and Wigner argued that since (global) charge-dependent phase transformations are physically unobservable, charge state superpositions are unphysical (`charge superselection rule'). This would resolve the apparent paradox in a trivial way. However, Aharonov and Susskind disputed this superselection rule: they distinguished between such global charge-dependent transformations, and transformations of the relative inter-charge phases of two particles, and showed that the latter \emph{could} in principle be observable! Finally, the paradox again disappears once we considers the `calibration' of the phase measured by the Aharonov-Susskind phase detectors, as well as the phase of the particle at its initial point. It turns out that such a detector can only distinguish between the relative phases of two paths if their (oriented) difference forms a loop around the flux. 
\end{abstract}

\vspace{2pc}
%\noindent{\it Keywords}: Aharonov Bohm Effect, Superselection rules, Topological phases 
%\submitto{\JPA}
\maketitle

\section{Introduction}

The Aharonov Bohm effect requires little introduction, nor is there
much need to convince the reader of its still inspiring present day
research, especially in this special volume celebrating its Golden
Jubilee. The Aharonov Susskind papers disputing the notion of superselection
rules in non-relativistic QM, on the other hand, although themselves
over 40 years old, and despite having introduced the very important
concept of quantum reference frames, have for decades remained mostly
unknown to all but quantum foundations specialists. Part of the reason,
at least, may have been the fact that they apparently deal exclusively
with the possibility of existence of coherent superpositions of states
with different electric charges or spin, which are admittedly esoteric
and seemingly irrelevant to most of the rest of physics. Nevertheless,
they have about a decade ago regained a much wider audience in the
context of apparent paradoxes related to the role played by coherent
superpositions of number states in quantum optics and the theory of
Bose-Einstein condensates. Ironically, the ubiquitous and innocent-looking
optical coherent state, which was one of the sources of inspiration
of AS, has now itself been claimed to be a ``convenient fiction''\cite{PhysRevA.55.3195},
and much of the theory utilizing it was in desperate need of being
reinterpreted\cite{RevModPhys.79.555}--a task accomplished by applying
to it the very same framework of quantum reference frames first invoked
by AS for the esoteric analogues mentioned earlier.

When a particle with electric charge $Q$ travels (slowly) along a
curve \emph{C} (i.e., its wavepacket is assumed to be small compared
to the length of the curve, and its centroid to move along the latter)
in the force-free region near a solenoid, it ostensibly picks up a
phase $\phi_{\text{AB}}^{\left[\boldsymbol{A}\right]}(C)\equiv Q\int_{C}\boldsymbol{A}\cdot d\boldsymbol{r}$,
where $\boldsymbol{A}$ is the vector potential generated by the current
in the solenoid. However, this formal expression is gauge-dependent
(hence the superscript label), and by gauge-invariance should be unobservable.
This unobservability also follows trivially from the fact that this
is just an overall phase of the state vector. One can get a similar
\emph{relative} phase if one considers the particle state to correspond
to a coherent superposition of localized packets following two paths
which share the same initial and final points. The relative phase
between the two branches of the wavefunction can then be measured
by interference at the final point. Astonishingly, this relative phase
turns out to be non-zero (the AB effect, of course), and is equal to
\begin{equation}\phi_{AB}(C)\equiv\phi_{\text{AB}}^{\left[\boldsymbol{A}\right]}(C_{1})-\phi_{\text{AB}}^{\left[\boldsymbol{A}\right]}(C_{2})=Q\oint_{C_{1}-C_{2}}\boldsymbol{A}\cdot d\boldsymbol{r},\end{equation}
where $C\equiv C_{1}-C_{2}$ is the directed difference of the two
curves--a closed loop. As we know, a simple calculation shows $\phi_{AB}(C)=nQ\Phi$,
where $\Phi$ is the magnetic flux threading the solenoid, and $n$
is the (signed) number of times $C$ winds around it (a purely topological
property). The RHS is manifestly gauge-invariant, which justifies
the dropping of the superscript on the LHS. 

What if you could have a coherent superposition of states with different
charges, say $Q$ and $0$? Suppose such a packet moved along $C$,
starting with a null relative phase. Then the same expression we had
before, $\phi_{\text{AB}}^{\left[\boldsymbol{A}\right]}(C)=Q\int_{C}\boldsymbol{A}\cdot d\boldsymbol{r}$,
would correspond to the \emph{relative} phase between the two components
after reaching the end of the curve! Are relative phases between different
charge states physically observable? Wick, Wightman and Wigner\cite{PhysRev.88.101}
argued they never are (for other reasons), and should be viewed as
notational redundancies of the formalism, on a par with the overall
phase. If the inter-charge phases at eiter end of $C$ is unobservable, as claimed, then so is their difference,
resolving our apparent paradox trivially. Ironically, Aharonov and Susskind\cite{PhysRev.155.1428} (AS)
argued that such phases \emph{are} in principle observable! In the
tradition of Yakir Aharonov\cite{aharonov2005quantum}, my mentor,
I will try to use this apparent paradox to gain a better understanding
of the charge superselection rule: to what extent can it can be violated
without violating gauge invariance.

\section{Superselection Rules and quantum reference frames}

\subsection{Are all Hermitian operators measurable?}

In Von Neumann's original axiomatic formulations of non-relativistic
quantum mechanics\cite{von1955mathematischen} (1932), the observables
of the theory were identified with the set of Hermitian operators
on the Hilbert space. The assumption that \emph{all} Hermitian operators
are observable implies that all \emph{rays} in Hilbert space are  physically
distinguishable. In other words, the only redundancy in the description
of states by normalized \emph{vectors} in Hilbert space was the overall
phase. Twenty years later, motivated by ambiguities in state description
in \emph{relativistic} quantum mechanics pointed out by Yang and Tiomno
\cite{PhysRev.79.495}, Wick, Wightman and Wigner\cite{PhysRev.88.101}(WWW)
restricted the class of observables in relativistic, \emph{as well
as non-relativistic} QM. They argued that superpositions of different
spin or charge eigenstates should be viewed as formal expressions
in which the phases between differing eigenspaces are operationally
meaningless, on a par with the aforementioned overall phase. Using
the term {}``selection rule'' synonymously with {}``conservation
law'', they coined the term {}``superselection rule'' for selection
rules which are mirrored in a fundamental limitation on \emph{measurements}
as well, i.e., measurements of phases between certain subspaces ({}``superselection
sectors'') are impossible: {}``We shall say that a superselection
rule operates between subspaces if there are neither spontaneous transitions
between their state vectors (i.e., if a selection rule operates between
them) and if, in addition to this, there are no measurable quantities
with finite matrix elements between their state vectors''. Linear
momentum is contrasted with intrinsic particle parity and with electrical
charge: the first has a selection rule, but no superselection rule,
the second has both and the third is postulated to also have both%
\footnote{If one adopts the point of view that quantum measurements consist
of interactions describable within the theory, as is now commonplace
(in defiance of Bohr's forebodings), then the very distinction between
selection and superselection rules seems to beg clarification (a possible
distinction was suggested by WWW in their reply to AS, \cite{PhysRevD.1.3267}).%
}. 

The strongest evidence for a superselection rule, according to WWW,
regards the unphysicality of the phase between the fermionic and bosonic
subspaces of the whole Hilbert space. They give a formal proof hinging
on the different effect the time inversion operator (introduced earlier
by Wigner), $T$, has on these two classes of states. Namely, applying
$T^{2}$ to a formal superposition of the two types of states will
result in a relative minus sign between them. Since $T^{2}$ should
have no observable effect, neither should the corresponding relative
phase (an illuminating analysis of this argument, as well as the AS
counter-argument appears in \cite{aharonov2005quantum}). A more circumstantial
argument (in their view) is given for the charge superselection rule,
based on the symmetries of the Hamiltonians then in use for field
theories for charged particles, which {}``in all cases {[}...{]}
is invariant against a simultaneous multiplication of all fields by
the same $e^{i\alpha}$. This property is known to be connected with
the principle of conservation of the total charge and represents a
very restricted type of gauge invariance... We are thus led to postulate
that: multiplication of the state vector $F$ by the operator $e^{i\alpha Q}$
produces no physically observable modification of the state of a system
of (mutually interacting) charged fields''.

\subsection{AS relativity: all selection rules on same footing, require reference
frames}

In 1967 these two superselection rules were challenged for the first
time. Aharonov and Susskind argued that, to begin with, the arguments
used by WWW for the parity- and charge-superselection rules applied
equally well to any and all selection rules, so if valid, they would
constitute a no-go theorem on measurements of linear and angular momentum!
They then proceeded to resolve the apparent paradox by analysing those
two examples in detail. Finally, they showed how an inter-charge phase
could be measured\cite{PhysRev.155.1428} in analogy to analogous
measurements in quantum optics (and similarly for the phase between
two different spin states\cite{PhysRev.158.1237}).

Aharonov and Susskind agreed that the relative phase induced between
different charge sectors by the global operation $e^{i\alpha\hat{Q}}$
is unobservable, when considering a closed system (such as the whole
universe), but argued that the operation $e^{i\alpha\hat{L}_{z}}$,
of rotating the entire universe about the $z$ axis, for example,
is just as physically meaningless. Only angles with respect to frames
defined by other physical objects are meaningful, and these are unchanged
when everything is rotated simultaneously. They consider a closed
system consisting of an electron and two large magnets. The two magnets
are assumed, without loss of generality to have zero total angular
momentum, and the electron to be in the $\sigma_{Z}=+1$ eigenstate
(in the frame defined by the previous assumption). The electron interacts
with the first magnet in such a way as to transform its spin state
to $\sqrt{1-r^{2}}|\sigma_{Z}=+1\rangle+re^{i\theta}|\sigma_{Z}=-1\rangle$.
This phase is evidently physically meaningful, and has a simple interpretation
in terms of the direction of polarization of the electron. This phase
can be measured using the second magnet, provided the relative orientation
of the two magnets is well defined. A simple analysis shows that the
two magnets can indeed be in an angular momentum eigenstate and have
an approximately well defined relative phase, i.e., the \emph{sum}
of the angular momenta and the \emph{difference} of the angles are
both well defined (this is analogous to the original EPR state being
a simultaneous eigenstate of $\boldsymbol{\hat{p}_{1}+\hat{p}_{2}}$
and $\boldsymbol{\hat{x}}_{1}-\boldsymbol{\hat{x}}_{2}$). The sharp
angular momentum implies that the angle $\theta$ is completely undefined
with respect to a reference frame external to the whole system, as
are all angular orientations of parts of the system. However, relative
angles are perfectly consistent with it. Similar arguments hold for
measurements of relative positions within a closed system, the whole
of which posses well defined momentum. 

To show how a charge-superposed nucleon state can be created, and
how the relative phase can be measured, AS describe a set-up analogous
to Ramsey interferometry of quantum optics. In Ramsey interferometry,
{}``two-level atoms'' are prepared in superpositions of ground and
excited states by passing through a cavity containing a coherent state
of photons, and the phase can be measured by passing them through
a second such cavity and measuring the final probability to be in
the excited state. The analogy is effected by replacing atoms by nucleons
(which can be in a superposition of isospin eigenstates: a proton
$|P\rangle$, and a neutron $|N\rangle$), and photons by charged
mesons (pions). Using the coherent state formalism introduced in quantum
optics by Glauber\cite{PhysRev.131.2766} a few years earlier, they
define coherent states of a charged meson field with mean charge $Q$
($\langle Q,\theta|\hat{Q}|Q,\theta\rangle=Q$) and phase $\theta:$ 

\begin{equation}
|\alpha=\sqrt{Q}e^{i\theta}\rangle=|Q,\theta\rangle=\sum_{n}\frac{Q^{n/2}}{\sqrt{n!}}e^{in\theta}|n\rangle
\end{equation}
(The Fock state $|n\rangle$ has $n$ units of charge: $\hat{Q}|n\rangle=n|n\rangle$).
A nucleon passing through a cavity containing such a state will experience
an effective interaction of the Jaynes-Cummings form:

\begin{equation}
H=g(t)\left(\sigma^{+}a^{-}+\sigma^{-}a^{+}\right),\end{equation}
where $g(t)$ is constant with value $g$ during time interval $\left[0,T\right]$
corresponding to the sojourn of the nucleon through the cavity (during
which the action of the free Hamiltonian can be neglected), and zero
outside it. 

A proton entering cavity $C1$ exits in a superposition state, which in the approximation $Q\gg1$, is given by:
\begin{equation}
|P\rangle|Q,\theta\rangle\longmapsto\cos\left(gTQ^{1/2}\right)|P\rangle+ie^{i\theta}\sin\left(gTQ^{1/2}\right)|N\rangle
\end{equation}
The absolute values of the coefficients of the charge eigenstates
clearly have operational meaning as probability amplitudes for the
respective states, but how can we measure the relative phase? Passing
the nucleon through a second cavity, $C2$, containing the mesonic
coherent state $|Q',\theta'\rangle$, we get a similar expression,
but now the \emph{absolute value} of each coefficient depends on $\theta-\theta'$.
Thus, the probability of the nucleon to exit the second cavity as
a proton depends on the relative phase of the fields in the two cavities.
In other words, measuring this probability is tantamount to measuring
the phase between the proton and neutron in the intermediate state,
relative to reference frame defined by $C2$. 

So far, so good, but we have resorted to using coherent mesonic states,
which themselves contain coherences between different charge eigenstates!
So far then, the logic seems to be circular (an analogous objection
regarding \emph{optical} coherent states would only be raised decades
later, as we shall note in the next section). However, AS note that
just as in the angular momentum case, we do not really need well defined
phases for each cavity separately, only a well defined relative phase
$\Delta\theta\equiv\theta-\theta'$, which \emph{is} consistent with
a well defined total charge $Q_{1}+Q_{2}$ for the two cavities. They
note that such a state is approximately given by:

\begin{equation}
|i\rangle=\int|Q,\theta_{1}\rangle|Q',\theta_{1}+\Delta\theta\rangle e^{i\left(Q+Q'\right)\theta_{1}}\end{equation}
(where the integration should be understood to be over $\theta_{1}$).
That this is indeed an eigenvector of $Q_{1}+Q_{2}$ is readily verified
by a direct calculation, but there is a slight subtlety with the interpretation
of this as having well defined relative phase. Not only is there no
phase operator (which is why we need to assume the \emph{Q}s are big),
but the values of the integrand evaluated at different values of $\theta_{1}$are
not mutually orthogonal (the coherent states, far from forming an
orthonormal basis, are actually overcomplete). This leaves a gap in
the proof due to the possibility that interference between different
terms would change the conclusion. AS also suggest a Gedankenexperiment
where a relative phase between the two cavities is established by
populating them jointly by passing $Q$ mesons through an appropriate
beam-splitter, such that the transmitted amplitude goes in one cavity
and the reflected one in the other. The relative phase would be determined
by the complex reflection and transmission coefficients (and the state
would also be an exact eigenstate of $Q_{1}+Q_{2}$). The optical
analogue of this set-up in this very context was rigorously analysed
much later by M$\text{ø}$lmer \cite{PhysRevA.55.3195}.

\subsection{WWW reply and aftermath: symmetry breaking primordial coherences
or ubiquitous reference frames?}

WWW \cite{PhysRevD.1.3267} replied to the AS papers by acknowledging
that \emph{given} a mesonic charged coherent state, and the physically
acceptable meson-nucleon interaction used by AS, an inter-charge sector
phase could indeed be measured. However, they claimed that the assumption
of its existence only begs the question: such coherence cannot be
created by the postulated interaction, and so it is required to have
pre-existent coherence to measure coherence (it takes one to know
one, so to speak), and in particular to create it by measurement %
\footnote{This in turn, seems would raise the question of the distinction between
selection and superselection. See previous footnote %
}. Since there is no evidence for what could be called {}``primordial
coherence'', there is no basis for assuming that such measurement
could be performed. The lack of superselection rules corresponding
to other selection rules is explained by the prior existence of just
such god-given coherences: {}``There is, in this regard a fundamental
difference between conserved quantities, such as linear and angular
momentum on the one hand, and electric (and baryonic) charge on the
other. We have naturally been given states which are superpositions
of states with different momenta; all more or less localized states
are of this nature...''. They do not address the issue of reference
frames. 

In an ironic turn of fate, much later, in a series of papers dealing with the status of the particle
number superselection rule in Bose-Einstein condensates\cite{PhysRevLett.76.161,PhysRevA.55.4330}
and optical systems such as the laser\cite{PhysRevA.55.3195}, the
measurability of optical phases (and their atomic-optical analogues)
has itself come under suspicion. In addressing these issues, the AS
reference frame concept has been called to the rescue of (the theoretical
explanation of) Ramsey interferometry itself and even the existence
of the paradigmatic optical coherent state itself, which inspired
it, thus coming full circle!

\section{Are reference frames for charge compatible with gauge invariance?}

We are now in a position to put together the different pieces of the
puzzle. Let us revisit the paradox described in the introduction,
and attempt to fill in the details in careful adherence to the AS
measurement procedure. AS tell us that we can, in principle, prepare
an initial state of the form $\frac{1}{\sqrt{2}}\left(|P\rangle+|N\rangle\right)$,
by letting it interact with a mesonic field in cavity \emph{$c1$}
at point $A$. The phase, $\varphi_{A}$, is defined with respect
to that of the cavity field. Now after travelling along curve $C$,
let it encounter a second cavity $c2$ at its other end, $B$. Its
phase relative to that of $c2$, $\varphi_{B}$, is again AS-measurable.
Now, 
\begin{equation}\varphi_{AB}\equiv\varphi_{B}-\varphi_{A}=Q\int_{C}\boldsymbol{A}\cdot d\boldsymbol{r}-\Delta\theta, \end{equation}
where $\Delta\theta$ is the relative phase between the two cavity
mesonic fields. Since preparation and measurement are closely related
in QM, this relative phase can be thought of as the calibration of
the cavities, viewed as measuring devices. However, this calibration
itself is gauge dependent. One way to see this is to consider two
similar cavities initially at $A$ containing identical coherent states,
$|Q,\theta\rangle$, and then letting one of them travel along a curve
$C'$ to point $B$. Described in a particular gauge, the states of
the fields in the stationary and mobile cavities at the end of this
process will be given by $|Q,\theta\rangle$ and $e^{i\phi_{\text{AB}}^{\left[\boldsymbol{A}\right]}(C')\hat{Q}}|Q,\theta\rangle=|Q,\theta+\phi_{\text{AB}}^{\left[\boldsymbol{A}\right]}(C')\rangle$,
respectively. Therefore, the $\varphi_{AB}=\phi_{\text{AB}}^{\left[\boldsymbol{A}\right]}(C)-\phi_{\text{AB}}^{\left[\boldsymbol{A}\right]}(C')$
which is equal to the topological (and gauge-independent) AB phase
$\phi_{\text{AB}}^{\left[\boldsymbol{A}\right]}(C-C')$. Thus, there
is some trivial freedom in calibration, but the only phase information
one obtains from this measurement procedure is the topological one.
In effect, the detector itself `closes the loop' (either around, or
outside the fluxon). The latter interpretation is analogous to a result
of Vaidman and Aharonov\cite{PhysRevLett.75.2063,PhysRevA.61.052108}
on the measurement of the relative phase between remote packets of
a single photon through its absorption by a pair of atoms and conversion
into a proper (two-particle) EPR-Bohm state.

\section{Summary}

In the AB effect, a charged particle travelling in the force-free region
outside of a solenoid, formally acquires a path- and gauge- dependent
phase, in a continuous fashion. However, what saves the day for gauge
invariance (or for the AB effect) is the fact that this phase is to
some extent an artifact of the notation. If our particle is the only
system treated as quantum, then this is a manifestly redundant overall
phase, except for situations where two amplitudes corresponding to
different sets of Feynman paths add up, which together can be considered
closing a loop, and thus is just the topological, non-local, situation,
which is gauge invariant. If, however, one allows \emph{coherent}
superpositions of different charge states, then the phase picked up
along an open trajectory is a \emph{local relative phase}, and hence
measurable! At first sight this seems to conflict with gauge invariance.
In fact, the AS phase, whatever its provenance is gauge dependent,
to begin with, just like the initial phase of the charged particle
in the regular AB effect. Since AS tell us that we should look at
phases defined relative to physical reference frames, we are led to
consider such frames for the two spatially separated endpoints of
the curve $C$. There is ambiguity in the relative phase of the latter,
due to gauge freedom. However, the phase difference between the initial
and final states \emph{relative to these frames} is gauge invariant
and depends only on the topological AB effect. We conclude that a reference 
frame for inter-charge phase can, in principle, be established locally in the Aharonov-Susskind sense, the relative
phase of two spatially separate phase standards is ambiguous, in accordance with gauge-invariance.

%\section*{Acknowledgements}
\ack
%It is my pleasure to acknowledge the helpful discussions with Lev Vaidman, Benni Reznik and Yakir Aharonov.
I wish to thank Lev Vaidman, Benni Reznik and Yakir Aharonov for helpful discussions. I acknowledge the support of the ISF, EC ( FET Open , MIDAS project) and DIP.

\section*{References}
\bibliographystyle{unsrt}
\bibliography{AB50Bib}

\end{document}